\documentclass[showpacs,preprintnumbers,twocolumn,aps,prl,reprint,superscriptaddress,nobibnotes,nofootinbib]{revtex4-1}

\usepackage{amsmath,amssymb,hyperref,times,graphicx,bm,bbold}
\usepackage{bibunits}
\def\<{\langle}
\def\>{\rangle}
\DeclareMathOperator{\Tr}{Tr}

\newcommand{\Xdag}[2]{\hat{\pmb{#1}}_{#2}^\dag}
\newcommand{\Xd}[2]{\hat{\pmb{#1}}_{#2}}

\newcommand{\adag}[1]{\Xdag{a}{#1}}
\newcommand{\ad}[1]{\Xd{a}{#1}}
\newcommand{\cdag}[1]{\Xdag{c}{#1}}
\newcommand{\cd}[1]{\Xd{c}{#1}}

\newcommand{\fd}[1]{\Xd{f}{#1}}
\newcommand{\gdag}[1]{\Xdag{g}{#1}}
\newcommand{\gd}[1]{\Xd{g}{#1}}
\newcommand{\nop}[1]{\Xd{n}{#1}}
\newcommand{\mop}[1]{\Xd{m}{#1}}

\newcommand{\id}{\mathbb{1}}

\newcommand{\EPL}{EPL}
\newcommand{\JSTAT}{J.~Stat.~Mech.}

\newcommand{\PRL}{Phys.~Rev.~Lett.}
\newcommand{\PRB}{Phys.~Rev. B}

\newcommand{\JHEP}{J.~High Energy Phys.}
\newcommand{\JMP}{J.~Mat.~Phys.}
\newcommand{\JPCM}{J.~Phys.: Condens.~Matter}
\newcommand{\JPAOLD}{J.~Phys.~A: Math.~Gen.}
\newcommand{\JPA}{J.~Phys.~A: Math.~Theor.}
\newcommand{\physrep}{Phys.~Rep.}

\newcommand{\PLB}{Phys.~Lett.~B}
\newcommand{\NJP}{New~J.~Phys.}

\newcommand{\SMtext}{See Supplemental Material for a discussion of the QMC sampling, a proof and generalization of Eq. (3), and a description of the procedure to explicitly evaluate Eq. (4)}

\begin{document}

\defaultbibliographystyle{apsrev4-1_custom}
\defaultbibliography{francesco,extra}
\begin{bibunit}
\title{Entanglement Hamiltonian of Interacting Fermionic Models}

\author{\firstname{Francesco} \surname{Parisen Toldin}}
\email{francesco.parisentoldin@physik.uni-wuerzburg.de}
\author{\firstname{Fakher F.} \surname{Assaad}}
\email{assaad@physik.uni-wuerzburg.de}
\affiliation{\mbox{Institut f\"ur Theoretische Physik und Astrophysik, Universit\"at W\"urzburg, Am Hubland, D-97074 W\"urzburg, Germany}}
\begin{abstract}
  Recent numerical advances in the field of strongly correlated electron systems allow the calculation of the entanglement spectrum and entropies  for interacting fermionic systems.   An explicit determination of the entanglement (modular) Hamiltonian has proven to be a considerably more difficult problem, and only a few results are available.
  We introduce a technique to directly determine the entanglement Hamiltonian of interacting fermionic models by means of auxiliary field quantum Monte Carlo simulations. We implement our method on the one-dimensional Hubbard chain partitioned into two segments and on the Hubbard two-leg ladder partitioned into two chains.   In both cases, we study the evolution of the entanglement Hamiltonian as a function of  the physical temperature. 
\end{abstract}

\maketitle

{\it Introduction.}---
The advent of quantum information techniques in the field of condensed matter physics has boosted a variety of new insights in old and new problems. In particular, recent years have witnessed a rapidly growing number of investigations of the quantum entanglement in strongly correlated many-body systems \cite{Laflorencie-16,PTA-18b}. The simplest approach is the so-called bipartite entanglement, where one divides a system into two parts, and a reduced density matrix describing one of the subsystems is obtained by tracing out the degrees of freedom of the other part. Arguably, the most studied quantities in this context are the entropies of the reduced density matrix, that is, the von Neumann and especially the Renyi entropies.  In the ground state, the entanglement entropies generically satisfy an area law; i.e., to leading order they are proportional to the area between the two subsystems \cite{ECP-10}. Among the many results, it is well established that in a 1+1 conformal field theory (CFT) corrections to the area law allow one to extract the central charge of a model \cite{CC-04}.

More information is contained in the entanglement Hamiltonian, also known as the modular Hamiltonian, which is defined as the negative logarithm of the reduced density matrix. Its spectrum, dubbed as the ``entanglement spectrum,'' has been shown to feature the edge physics of topologically ordered phases such as the fractional quantum Hall state \cite{LH-08}  as well as of symmetry-protected topological states of matter   \cite{PTA-18b,Fidkowski-10,TZV-10,ALPT-13,Assaad-15}. The entanglement Hamiltonian also plays a central role in the first law of entanglement \cite{BCHM-13}. Beside the entanglement spectrum and the associated eigenvectors, the knowledge of the entanglement Hamiltonian opens the possibility of characterizing the reduced density matrix as a thermal state. Furthermore, the expectation value of the entanglement Hamiltonian equals the von Neumann entanglement entropy, a key quantity which is generically not accessible in numerical simulations of interacting models.
Perhaps not surprisingly, compared to the computation of entanglement entropies, an explicit determination of the entanglement Hamiltonian has proven to be a considerably more difficult problem, and only a few solvable results are available. Aside from limiting cases, such as in the absence of interactions between the two subsystems, or the high-temperature limit, where the entanglement Hamiltonian can be easily determined, a particularly important result concerns a relativistic field theory in flat $d$-dimensional Minkowski space. For a bipartition of the space into two semi-infinite subsystems with no corners, translationally invariant along $d-1$ dimensions, the entanglement Hamiltonian is given by an integral of the energy-momentum tensor, with a weight proportional to the distance $x$ from the boundary, leading to the Bisognano-Wichmann (BW) form of the entanglement Hamiltonian \cite{BW-75,BW-76}. In the presence of additional conformal symmetry, a mapping of the semi-infinite space to a ball allows one again to express the entanglement Hamiltonian as an integral of the energy-momentum tensor, with a space-dependent weight \cite{CHM-11}. Reference \cite{CT-16} provides a recent review of the cases in $1+1$ CFT where the entanglement Hamiltonian is obtained as a weighted integral of the energy-momentum tensor.

Concerning condensed matter models on a lattice, the entanglement Hamiltonian is exactly known only in a few cases in one dimension and for a semi-infinite line subsystem: the noncritical transverse-field Ising model and the $XXZ$ model in the massive phase \cite{IT-87,PKL-99}.
Even in the deceptively simple case of a free (nonrelativistic) fermionic chain, the explicit computation of the entanglement Hamiltonian for a segment proved to be a rather difficult task. Although for a free fermionic system an exact formula for the entanglement Hamiltonian is known \cite{Peschel-04}, its explicit calculation for a finite segment embedded in a chain has eluded an analytical treatment so far.
All lattice models mentioned above share the property of being described by a CFT in the low-energy limit; hence, the entanglement Hamiltonian should attain the BW form, as indeed confirmed by the exact determination for the Ising and $XXZ$ models. Nevertheless, the entanglement Hamiltonian of the free fermionic chain model contains intriguing corrections to the CFT prediction which, remarkably, persist even in the limit of a long segment \cite{EP-17}.
In this context, recent studies have provided numerical evidence in support of a lattice-discretized BW form of the entanglement Hamiltonian for various models in both one and two dimensions \cite{KKTH-16,DVZ-18,KLC-18,ZHH-18,GMSCD-18}.

In this Letter we introduce a numerically exact quantum Monte Carlo (QMC) method which allows one to determine the entanglement Hamiltonian of interacting model of fermions. The method is applied to the Hubbard chain and to the two-leg Hubbard ladder, where we compute the one- and two-body terms of the entanglement Hamiltonian as a  function of the temperature. 

{\it Method.}---
The method presented here is based on QMC simulations using the auxiliary field algorithm \cite{AF_notes,BSS-81,WSSLGS-89}, whose basic formulation is reported in the following. The Hamiltonian of the system $\hat{H}$ is separated into a sum of a free part $\hat{T}$ (containing, e.g., hopping terms) and an interaction part $\hat{V}$ (e.g., a Hubbard repulsion).  At finite inverse temperature $\beta$, one introduces a Trotter decomposition of the density matrix operator $\exp(-\beta \hat{H})$; in the models considered here, we found important to choose a symmetric decomposition $\exp(-\beta\hat{H}) =[\exp(-\Delta\tau\hat{T}/2)\exp(-\Delta\tau\hat{V})\exp(-\Delta\tau\hat{T}/2)]^N+O(\Delta\tau^2)$, with $\beta=N\Delta\tau$ thereby ensuring the Hermiticity of the imaginary time propagation.    Then, the interaction term is decoupled via a Hubbard-Stratonovich (HS) decomposition, introducing discrete HS fields $\{s\}$. The QMC simulation consists in a stochastic sampling of the probability distribution $P(\{s\})$ associated with the HS fields. The ALF package provides a framework to program auxiliary field QMC simulations \cite{ALF}. The introduction of the HS transformation results in a free fermionic system in the HS fields $\{s\}$. For such a system, the reduced density matrix associated with a subpartition $A$ of the Hilbert space can be written exactly in terms of the Green's functions of the model, restricted to the subsystem $A$ \cite{Peschel-03}. One then arrives to the following expression for the reduced density matrix $\hat{\rho}_A$ \cite{Grover-13}:
\begin{equation}
  \begin{split}
  &\hat{\rho}_A = \int d\{s\} P(\{s\}) \det\left[\id - G_A(\{s\})\right] e^{-\adag{i}h_{ij}(\{s\})\ad{j}},\\
    &h(\{s\}) = \log\left\{ \left[G_A(\{s\})^T\right]^{-1} -\id \right\},
  \end{split}
  \label{rhoAF}
\end{equation}
where $\adag{i}$ and $\ad{i}$ are the fermionic creation and annihilation operators, respectively, in the subsystem $A$ and $i$ and $j$ are superindices labeling the possible states in $A$; here and in the following, we assume an implicit summation over repeated indices. The Green's function matrix $G_A(\{s\})_{ij}\equiv \<\adag{i}\ad{j}\>$ restricted to $A$, and at a given configuration of $\{s\}$, is readily accessible in the auxiliary field algorithm, and it is computed at a fixed imaginary-time slice. Equation (\ref{rhoAF}) has been exploited to compute the Renyi entropies \cite{Grover-13,ALPT-13,DP-15,DP-16}. Alternatively, Renyi entropies can be computed by means of the replica trick, in fermionic \cite{BT-14,WT-14,Assaad-15,BT-16} and bosonic \cite{IHM-11} as well as spin systems \cite{HGKM-10,HR-12}.

Equation (\ref{rhoAF}) suggests to introduce a new measure $\widetilde{P}(\{s\}) \propto P(\{s\})\det\left[\id - G_A(\{s\})\right]$, such that $\hat{\rho}_A$ is obtained as an expectation value over the measure $\widetilde{P}(\{s\})$:
\begin{equation}
  \hat{\rho}_A \propto \< e^{-\adag{i} h_{ij}(\{s\})\ad{j}}\>_{\widetilde{P}}.
  \label{rhoPtilde}
\end{equation}
As discussed in the Supplemental Material \cite{SM}, for the models considered here it can be proven that $P(\{s\})$ as well as the determinant $\det\left[\id - G_A(\{s\})\right]$ are positive; hence, $\widetilde{P}(\{s\})$ can be sampled by QMC simulations without a sign problem. Furthermore, as proven in Ref.~\cite{SM}, the exponential on the right-hand side of Eq.~(\ref{rhoPtilde}) admits an expansion in normal-ordered many-body operators:
\begin{equation}
\begin{split}
&e^{-\adag{i} h_{ij}(\{s\})\ad{j}} = 1 + \adag{i}\left(e^{-h(\{s\})}-\id\right)_{ij}\ad{j} \\
&+ \frac{1}{2}\adag{i} \adag{k} \left(e^{-h(\{s\})}-\id\right)_{ij}\left(e^{-h(\{s\})}-\id\right)_{kl}\ad{l} \ad{j} + \cdots.
\end{split}
\label{rhoexpansion}
\end{equation}
By inserting Eq.~(\ref{rhoexpansion}) in Eq.~(\ref{rhoPtilde}), we obtain an expansion of $\hat{\rho}_A$ as a sum of many-body operators, whose coefficients can be sampled with a QMC simulation. This gives us an unbiased QMC determination of $\hat{\rho}_A$.

In order to compute the entanglement Hamiltonian $\hat{H}_E$, we first calculate the matrix elements $M$ of $\hat{\rho}_A$. The matrix $N\equiv -\log(M)$ represents, by definition, the matrix elements of $\hat{H}_E$. The entanglement Hamiltonian is then obtained by determining the many-body operator whose matrix elements are $N$. As for $\hat{\rho}_A$, we expand $\hat{H}_E$ as a sum of normal-ordered many-body operators:
 \begin{equation}
   \hat{H}_E=-\log(\hat{\rho}_A)=\text{const}-\adag{i}t_{ij}\ad{j}+\adag{i}\adag{k}U_{ijkl}\ad{l}\ad{j}+\cdots.
   \label{HEexpansion}
 \end{equation}
 Crucially, it is possible to prove that, in order to compute $\hat{H}_E$ up to the two-body term $\adag{i}\adag{k}U_{ijkl}\ad{l}\ad{j}$, it is sufficient to truncate the sampling of $\hat{\rho}_A$ to the two-body term, as done on the right-hand side of Eq.~(\ref{rhoexpansion}). Under this condition, the computational cost for sampling $\hat{\rho}_A$ and determining $\hat{H}_E$ is only polynomial in the size of the subsystem $A$.
As discussed in Ref.~\cite{SM}, the expansion of Eqs.~(\ref{rhoexpansion}) and (\ref{HEexpansion}) can be extended to any order in a sum of normal-ordered many-body operators, whose coefficients could be, in principle, sampled as to determine $\hat{\rho}_A$ and $\hat{H}_E$ beyond the two-body terms.
More technical details on this step of the algorithm, implemented using the TRIQS \cite{triqs} and \textsc{Armadillo} \cite{armadillo,armadillo18} libraries,
are reported in Ref.~\cite{SM}.

\begin{figure}[b]
  \centering
  \vspace{0.8em}
    \includegraphics[width=0.9\linewidth]{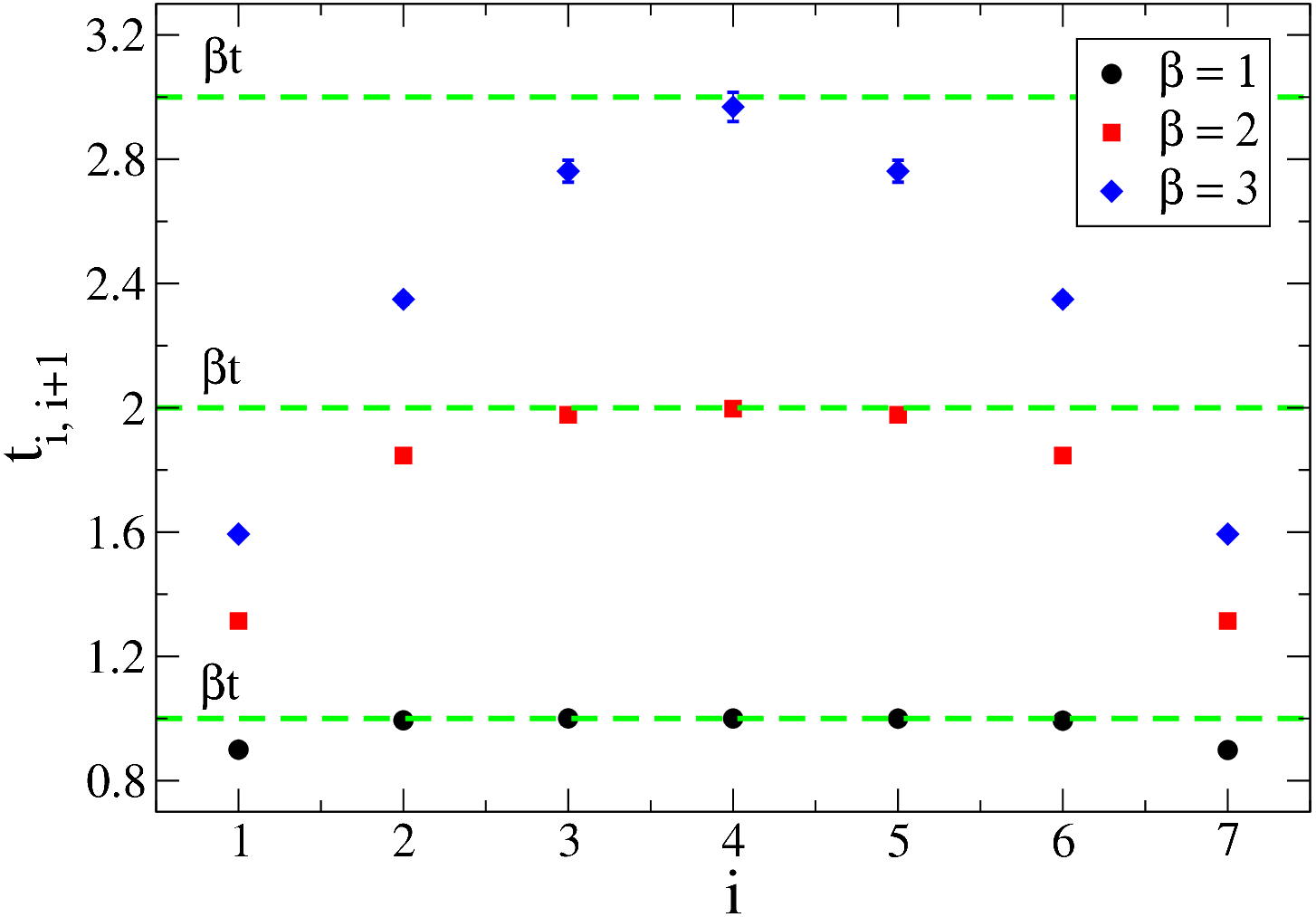}
  \caption{Nearest-neighbor hopping terms of the entanglement Hamiltonian for a segment of $L_a=8$ sites in a Hubbard chain of total length $L=32$, and parameters $t=1$ and $U=4$, as a function of the temperature. The dashed line indicates the expected values far from the boundary; see the main text.}
  \label{tchain}
\end{figure}

{\it Results.}---
We have applied the method outlined above to the Hubbard chain and the Hubbard two-leg ladder  at half-filling. The Hamiltonian for a Hubbard chain of length $L$ is
\begin{equation}
  \begin{split}
    \hat{H}=&-t\sum_{i=1,\sigma}^L \cdag{i,\sigma} \cd{i+1,\sigma} + \cdag{i+1,\sigma} \cd{i,\sigma} \\
    &+ U \sum_{i=1}^L \Big(\hat{\pmb{n}}_{i,\uparrow}-\frac{1}{2}\Big) \Big(\hat{\pmb{n}}_{i,\downarrow}-\frac{1}{2}\Big), \quad \sigma=\uparrow,\downarrow,
  \end{split}
  \label{Hchain}
\end{equation}
where by imposing periodic boundary conditions we identify the lattice site $L+1$ with $1$. For this model, we cut a subsystem $A$ consisting in a segment of length $L_a$ and compute the one-body term $t_{ij}$ defined in Eq.~(\ref{HEexpansion}). In Fig.~\ref{tchain}, we show the resulting hopping terms $t_{i,i+1}$ between nearest-neighbor lattice sites $i$ and $i+1$, as a function of $i$ and for three inverse temperatures $\beta=1$, $2$, and $3$. At a high temperature $\beta=1$, we find that $t_{i,i+1}$ attains the value of $1$ for all lattice sites except those next to the boundary. In fact, if the entanglement between $A$ and $B$ is locally restricted to a region close to the boundaries, we expect that, away from such a region, the subsystem $A$ is substantially independent of $B$, and, hence, locally, the entanglement Hamiltonian should match $\beta \hat{H}_A$, with $\hat{H}_A$ the Hamiltonian of the model, restricted to $A$. Accordingly, we observe a plateau with $t_{i,i+1}\simeq \beta t=\beta$ in the central part of the plots in Fig.~\ref{tchain}, whose extension shrinks as the temperature is lowered and the entanglement grows. For $\beta=3$ only for a single site in the middle we find $t_{i,i+1}\simeq \beta$, whereas close to the boundaries we observe an approximately linear dependence of $t_{i,i+1}$ on $i$, which grows (respectively, decreases) when close to the left (respectively, right) boundary. Such a behavior resembles qualitatively the case of a CFT \cite{CT-16}. For the Hamiltonian parameters considered, the other hopping terms in $\hat{H}_E$ are negligible.   For reasons expanded upon in Ref.~\cite{SM}, it is technically hard, for this specific model, to  reach lower temperatures  and especially to investigate temperature  scales  below which the magnetic correlation length is substantial.
Nevertheless, as a comparison in order to reproduce the results of Fig.~\ref{tchain} by exact diagonalization (ED) techniques, one would need a full-spectrum diagonalization of a Hubbard chain with size $L=32$, a task far beyond current ED capabilities.

In contrast,  for the Hubbard model on a two-leg ladder, we were able to reach {\it low}   temperatures, approaching the ground state.  The Hamiltonian is defined as
\begin{equation}
  \begin{split}
    &\hat{H}=-t\sum_{\substack{i,\sigma \\ O=A,B}} \cdag{i,O,\sigma} \cd{i+1,O,\sigma} + \cdag{i+1,O,\sigma} \cd{i,\sigma} \\
    &-t_\perp\sum_{i,\sigma} \cdag{i,A,\sigma} \cd{i,B,\sigma} + \cdag{i,B,\sigma} \cd{i,A,\sigma}\\
    &+ U \sum_{i,O=A,B} \Big(\hat{\pmb{n}}_{i,O,\uparrow}-\frac{1}{2}\Big) \Big(\hat{\pmb{n}}_{i,O,\downarrow}-\frac{1}{2}\Big),
  \end{split}
  \label{Hladder}
\end{equation}
where $t$ and $t_\perp$ indicate the intra- and interleg hopping constants, respectively, and $A$ and $B$ label the two legs. For this geometry, we trace out one leg and obtain a translationally invariant entanglement Hamiltonian for a single leg, i.e., defined on a chain geometry. At half filling, the ground state of the model consists of a single fully  gapped phase \cite{NWS-96,WOHB-01}. The charge gap $\Delta_C$ and the spin gap $\Delta_S$, with $\Delta_C > \Delta_S$, are monotonically increasing with $t_\perp$ and $U$. Gapped systems exhibit, as a function of the linear size, a fast approach to the thermodynamic limit \cite{Neuberger-89,WAPT-17}.

\begin{figure}[b]
  \centering
  \vspace{0.5em}
  \includegraphics[width=0.9\linewidth]{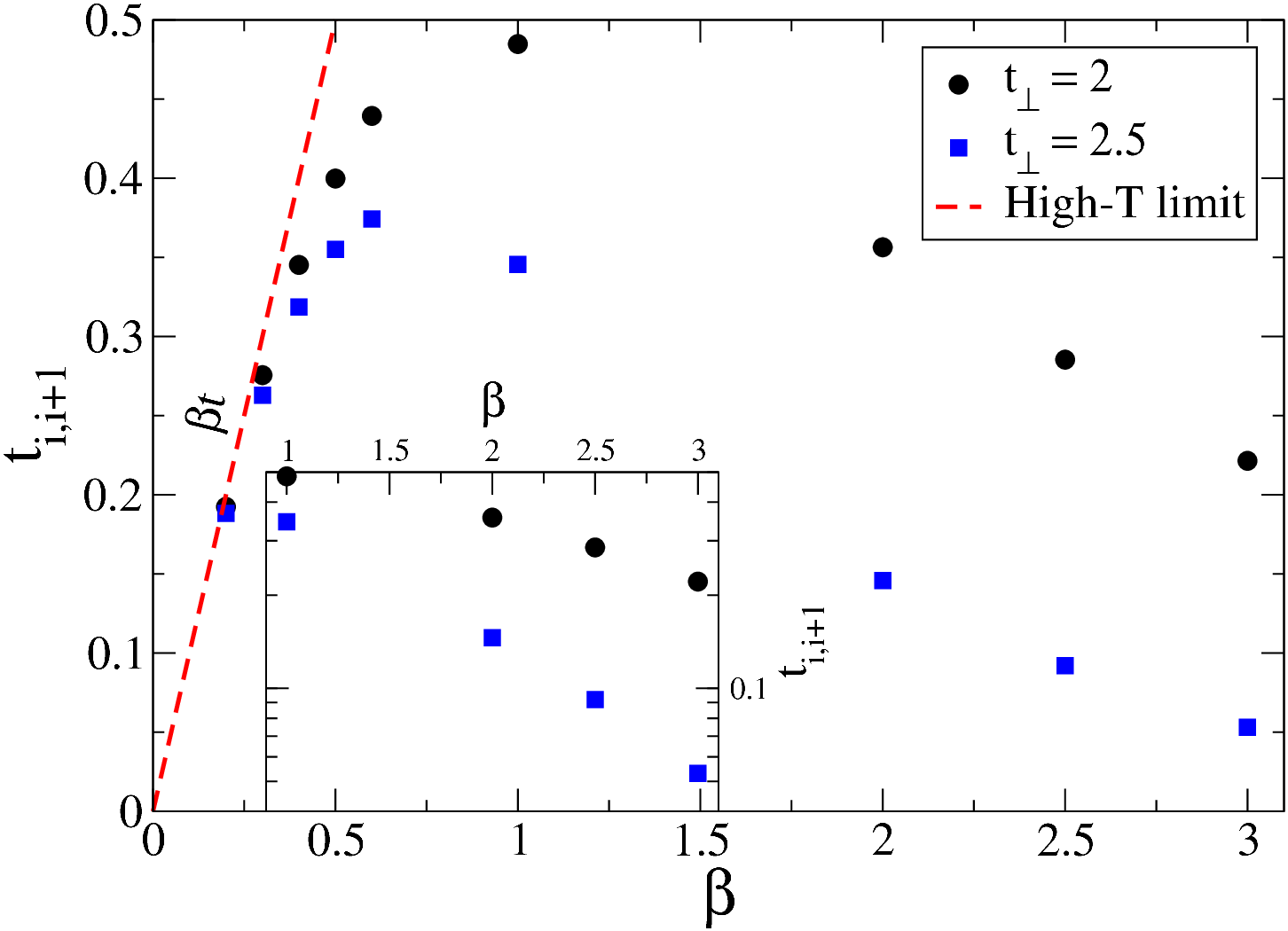}
  \caption{The same as Fig.~\ref{tchain}, for a Hubbard two-leg ladder of linear length $L=8$, with fixed parameters $t=1$ and $U=4$, as a function of the temperature for two values of $t_\perp$. The dashed line indicates the expected high-temperature limit $t_{i,i+1}\simeq\beta t$. The inset shows a magnification of the plot for $\beta\ge 1$ in a semilogarithmic scale.}
  \label{tladder}
\end{figure}

Figure \ref{tladder} illustrates the temperature dependence of the nearest-neighbor hopping term $t_{i,i+1}$ in $\hat{H}_E$ for $t_\perp=2$ and $2.5$ and fixed coupling constants $t=1$ and $U=4$. At high temperatures, $t_{i,i+1}$ grows linearly with $\beta$, $t_{i,i+1}\simeq\beta t$, in agreement with the theoretical expectation $\hat{H}_E=\text{const} + \beta \hat{H}_A + O(\beta^2)$, $\beta\rightarrow 0$, which follows easily by Taylor expanding the density matrix $\rho\sim\exp(-\beta\hat{H})$ to the lowest order in $\beta$. Upon decreasing the temperature, one eventually crosses the charge and spin gaps, leading to a suppression of the charge fluctuations. The entanglement Hamiltonian reflects this physics, showing a nonmonotonic temperature dependence of $t_{i,i+1}$, which starts to decrease for large enough values of $\beta$. The value of $\beta$ at which $t_{i,i+1}$ stops to grow decreases upon increasing $t_\perp$, because the gaps increase with $t_\perp$ \cite{NWS-96,WOHB-01}. Figure \ref{tladder} confirms this observation. Furthermore, a semilog plot of the data shown in the inset in Fig.~\ref{tladder} supports an exponential suppression of the hopping constants $t_{i,i+1}$ for $\beta\rightarrow\infty$. We notice that the charge gaps and spin gaps are $\Delta_C\approx 1.6$ and $\Delta_S\approx 0.6$ for $t_\perp=2$ and $\Delta_C\approx 2.1$ and $\Delta_S\approx 1.3$ for $t_\perp=2.5$ \cite{NWS-96}, respectively; hence, the data in Fig.~\ref{tladder} for the largest values of $\beta$ are well below the gaps and essentially approach the ground state of the model.
Hopping terms $t_{i,i+x}$ at distances $x>1$ are negligible compared to the nearest-neighbor one $t_{i,i+1}$.

\begin{figure}[t]
  \centering
  \includegraphics[width=0.9\linewidth]{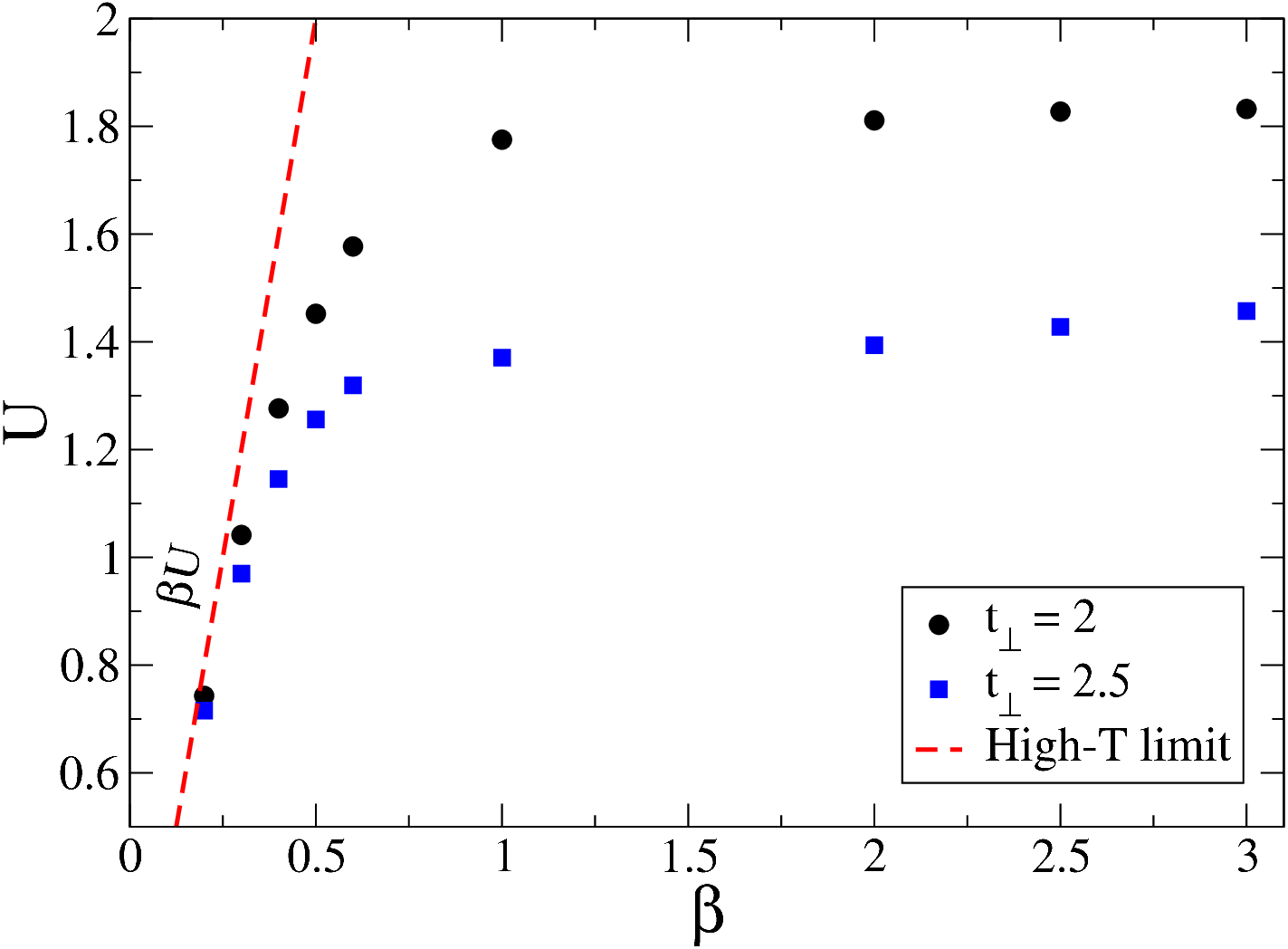}
  \caption{The same as Fig.~\ref{tladder} for the on-site Hubbard repulsion $U$.}
  \label{Uladder}
\end{figure}

\begin{figure}[t]
  \centering
  \includegraphics[width=0.9\linewidth]{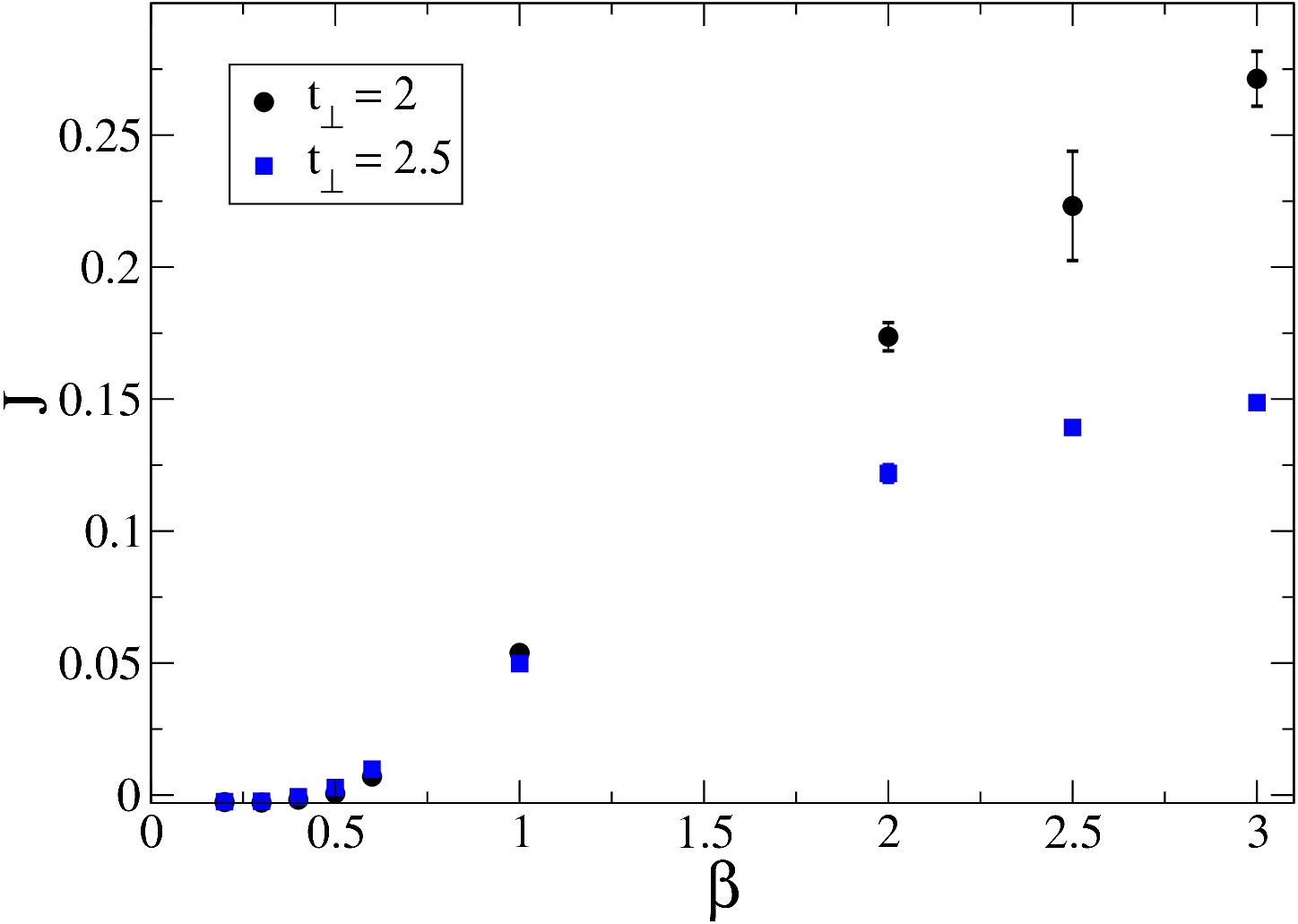}
  \caption{The same as Fig.~\ref{tladder} for the nearest-neighbor spin-spin interaction $J$.}
  \label{Jladder}
\end{figure}

\begin{figure}[b]
  \centering
  \vspace{0.5em}
  \includegraphics[width=0.9\linewidth]{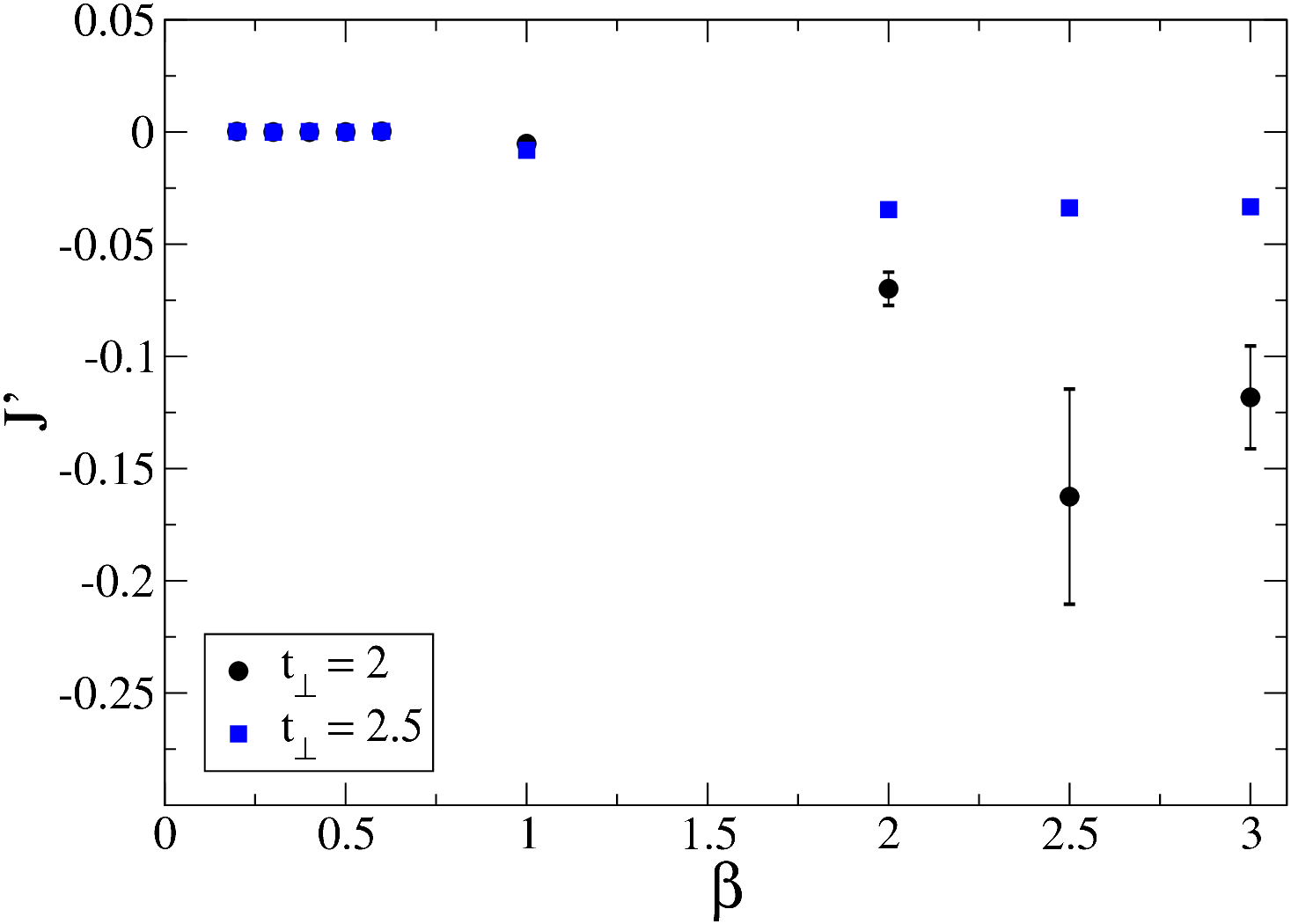}
  \caption{The same as Fig.~\ref{tladder} for the next-nearest-neighbor spin-spin interaction $J'$.}
  \label{Jpladder}
\end{figure}

For this model, we are able to compute all two-body terms in $\hat{H}_E$. In Fig.~\ref{Uladder}, we show the on-site Hubbard repulsion term $U$. As for the hopping term, it exhibits the expected linear increase with $\beta$ for high temperatures. However, upon crossing the gaps, $U$ saturates to a $t_\perp$-dependent value. The entanglement Hamiltonian contains also interaction terms which are absent in Eq.~(\ref{Hladder}), such as a nearest-neighbor antiferromagnetic spin-spin interaction $J\vec{\pmb{S}}_i\vec{\pmb{S}}_{i+1}$ and a next-nearest-neighbor ferromagnetic interaction $J'\vec{\pmb{S}}_i\vec{\pmb{S}}_{i+2}$, displayed in Figs.~\ref{Jladder} and \ref{Jpladder}. Both $J$ and $J'$ vanish at high temperatures, as expected, and grow only when the temperature is below the gaps of the model.
Additional two-body terms such as particle-particle interactions $V_{ij}\hat{\pmb{n}}_{i}\hat{\pmb{n}}_{i}$ and spin-spin interactions at distances larger than $2$ are effectively negligible compared to those shown in Figs.~\ref{Uladder}--\ref{Jpladder}.
All in all, the entanglement Hamiltonian exhibits a remarkable crossover between a Hubbard-like Hamiltonian at high temperatures, where $\hat{H}_E\simeq\beta \hat{H}_A + \text{const}$ to a Heisenberg-like Hamiltonian at low temperatures, where $U\gg t$ and additional nonfrustrating spin-spin interactions $J$ and $J'$ enforce an antiferromagnetic order. Such a behavior is analog to what is found in the two-leg Heisenberg model, where, for antiferromagnetic interchain and intra-chain couplings, the entanglement spectrum matches the one for a Heisenberg chain \cite{Poilblanc-10,CPSV-11}, as confirmed also by perturbative calculations showing that for strong rung coupling the entanglement Hamiltonian is approximately proportional to the restriction of the Hamiltonian to a single leg \cite{PC-11,LS-12,SL-12}; similar results have been obtained, e. g., in the case of free fermions \cite{PC-11}, bilayer quantum Hall systems \cite{Schliemann-11}, and Hofstadter bilayers \cite{Schliemann-13,*Schliemann-13_erratum} (see also Ref.~\cite{Schliemann-14} and references therein).
We notice that our results outperform ED, because a full spectrum diagonalization of a Hubbard model, needed to reproduce Figs.~\ref{tladder}--\ref{Jpladder}, is currently feasible for lattices
with $N\lesssim 12$ sites \cite{AL-communication}, corresponding to a $L=6$ two-leg ladder.

{\it Summary.}---
In this Letter, we present a general framework for computing the reduced density matrix and the entanglement Hamiltonian of an interacting fermionic model. The method is formulated within the auxiliary field QMC method and allows one to unbiasly determine the reduced density matrix and the entanglement Hamiltonian as a series of normal-ordered many-body operators. The method is applied to the Hubbard chain and two-leg models, where we present the first numerically exact determination of the one-body and two-body terms of the entanglement Hamiltonian. The results clearly show the increase of correlations and entanglement upon lowering the temperature, and for the two-leg model a change in the physical behavior of the model upon crossing the gaps, with the emergence of qualitatively different interactions in the entanglement Hamiltonian. Our results outperform current ED techniques; in fact, even if the ground state or the full spectrum is obtained by ED, the determination of the entanglement Hamiltonian requires the highly numerically unstable computation of the logarithm of the reduced density matrix. Thus, we expect our findings to provide a benchmark for future studies. The generality of the method described here paves the way for future investigations of interacting models, where the knowledge of the entanglement Hamiltonian may provide new useful insights. In fact, almost all of the entanglement measures can be, in principle, obtained from the entanglement Hamiltonian. Its determination with the present method can then allow one to compute key quantities otherwise inaccessible to numerical simulations, such as the entanglement negativity and the von Neumann entanglement entropy, which is simply equal to the expectation value of the entanglement Hamiltonian.
The method lends itself to study the reduced density matrix for a subsystem $A$ embedded into a potentially large system $B$. Thus, one may investigate the extension and space dependence of entanglement by, e.g., considering a possibly small, spatially disconnected, subsystem $A$.

F.P.T. is grateful to Igor Krivenko for technical support on the TRIQS library \cite{triqs} and to Didier Poilblanc and German Sierra Rodero for useful discussions. We thank Andreas L\"auchli and John Schliemann for useful communications. F.P.T.  thanks the German Research Foundation (DFG)  through Grant No. AS120/13-1   of the FOR 1807.   F.F.A.  thanks  the DFG for financial support through the SFB 1170 ToCoTronics. We acknowledge the computing time granted by the John von Neumann Institute for Computing (NIC) and provided on the supercomputer JURECA \cite{Jureca16} at the J\"ulich Supercomputing Center.

\putbib
\end{bibunit}
\clearpage

\begin{bibunit}
\section{Supplemental Material}

\subsection{QMC simulations}
\label{sm:positivity}
In this section we discuss some technical details of the QMC simulations. First we prove that the measure $\widetilde{P}(\{s\}) \propto P(\{s\})\det\left(\id - G_A(\{s\})\right)$ used in Eq.~(\ref{rhoPtilde}) is positive, hence amenable to Monte Carlo (MC) sampling. In the case of a repulsive Hubbard interaction
\begin{equation}
  \hat{V} = U\left(\hat{\pmb{n}}_\uparrow-\frac{1}{2}\right) \left(\hat{\pmb{n}}_\downarrow-\frac{1}{2}\right) = -\frac{U}{2}\left(\hat{\pmb{n}}_\downarrow-\hat{\pmb{n}}_\uparrow \right)^2+\frac{U}{4},
  \label{positiveU}
\end{equation}
up to an irrelevant constant, the HS decomposition can written as \cite{AF_notes}
\begin{equation}
  \begin{split}
    \exp&\{\Delta\tau U\left(\hat{\pmb{n}}_\downarrow-\hat{\pmb{n}}_\uparrow \right)^2/2\}\\
    &= \frac{1}{4} \sum_{ s = \pm 1, \pm 2}  \gamma(s) e^{ \sqrt{\Delta \tau U/2 } \eta(s)\left(\hat{\pmb{n}}_\downarrow-\hat{\pmb{n}}_\uparrow \right) }
                + {\cal O} (\Delta \tau ^4)\;,
  \end{split}
  \label{HSrepulsive}
\end{equation}
with $ \gamma(\pm 1)  = 1 + \sqrt{6}/3$, $\eta(\pm 1 ) = \pm \sqrt{\smash[b]{2 (3 - \sqrt{6} )}} $, and 
$  \gamma(\pm 2) = 1 - \sqrt{6}/3$, $\eta(\pm 2 ) = \pm \sqrt{\smash[b]{2 (3 + \sqrt{6} )}}$. This choice of the HS transformation does not introduce a sign problem, hence $P(\{s\})>0$ \cite{AF_notes}. Due to the conservation of the $\hat{S}^z-$symmetry on every configuration of the HS fields $\{s\}$, the determinant $\det\left(\id - G_A(\{s\})\right)$ appearing in Eq.~(\ref{rhoAF}) factorizes into
\begin{equation}
  \begin{split}
    \det&\left(\id - G_A(\{s\})\right)\\
    &=\det\left(\id - G_A^\uparrow(\{s\})\right) \det\left(\id - G_A^\downarrow(\{s\})\right),
  \end{split}
  \label{factorization}
\end{equation}
where $G_A^\sigma(\{s\})$ indicates the Green's function matrix in the $\sigma-$sector, $\sigma=\uparrow$, $\downarrow$. Being the HS decomposition of Eq.~(\ref{HSrepulsive}) real, the involved Green's function matrices are real. Furthermore, $G_A^\downarrow(\{s\})$ can be obtained by applying the antiunitary particle-hole transformation to $G_A^\uparrow(\{s\})$. By the usual Kramers degeneracy theorem, it follows that to each eigenvalue $\lambda$ of $G_A^\uparrow(\{s\})$ corresponds a complex-conjugate eigenvalue $\lambda^*$ of $G_A^\downarrow(\{s\})$, such that the product on the right-hand side of Eq.~(\ref{factorization}) is positive. We notice that the positiveness of $\widetilde{P}(\{s\})$ can also be easily proved in the case of an attractive Hubbard interaction, by using a real HS decomposition that preserves the full $SU(2)$ symmetry \cite{AF_notes}. In this case the positivity immediately follows from $\det\left(\id - G_A(\{s\})\right) = \det\left(\id - G_A^\uparrow(\{s\})\right)^2$.
In both cases, a key ingredient for the absence of a sign problem lies in the fact that the subsystem $A$ contains both spins degrees of freedom for each lattice site.

We observed two main limitations to the method presented in this letter. The first one is related to the stability of the MC sampling of the measure of $\widetilde{P}(\{s\})$. In the auxiliary field QMC method, the probability measure $P(\{s\})$ is, up to a normalization constant,
\begin{equation}
  \begin{split}
    &P(\{s\}) \propto \alpha(\{s\}) \det\left(\id-G(\{s\})\right)^{-1},\\
    &\alpha(\{s\})\equiv \prod_i\gamma(s_i),
  \end{split}
  \label{P}
\end{equation}
where $G(\{s\})$ is the full Green's function matrix at a given HS configuration $\{s\}$ and fixed imaginary-time slice. In Eq.~(\ref{P}), the term $\alpha(\{s\})$, inessential for the present discussion, is the product of $\gamma(s_i)$ factors introduced in Eq.~(\ref{HSrepulsive}), each depending on the HS field $s_i$ introduced to decouple the Hubbard interaction at the site $i$. It follows from Eq.~(\ref{P}) that configurations with large values of the Green's function are effectively suppressed in a stochastic sampling of $P(\{s\})$. On the other hand, when considering the modified measure $\widetilde{P}(\{s\})$
\begin{equation}
  \widetilde{P}(\{s\}) \propto \alpha(\{s\}) \det\left(\id-G(\{s\})\right)^{-1}\det\left(\id - G_A(\{s\})\right),
  \label{Ptilde}
\end{equation}
the additional factor $\det\left(\id - G_A(\{s\})\right)$ can reduce such a suppression, leading to larger fluctuations in the values of the Green's function during a MC sampling of $\widetilde{P}(\{s\})$. This effect becomes more pronounced on lowering the temperature, potentially hindering a reliable sampling. A peculiarity of the auxiliary field algorithm is the so-called ``fast update'': in a Metropolis sampling scheme, when updating the HS configuration at a given imaginary time slice, the acceptance probability depends only on the Green's function at the same time slice. Moreover, when a move is accepted, a fast update of the same Green's function can be implemented via the Sherman-Morrison formula \cite{AF_notes}. Thus, fast updates avoid computationally expensive calculations of the determinant in Eq.~(\ref{P}). However, in a finite-temperature simulation they give rise to a numerical error which cumulates over successive applications of Sherman-Morrison formula. To numerically stabilize the algorithm, one introduces a systematic recalculation of the Green's function matrix, whose value after the recalculation is compared to the one obtained via the fast updates, as to ensure numerical correctness. We defer to Ref.~\cite{AF_notes} for a discussion on the numerical stabilization of finite-temperature auxiliary field simulations. The fast update method can also be formulated in the case of the measure $\widetilde{P}(\{s\})$ of Eq.~(\ref{Ptilde}). On the other hand, the appearance of larger fluctuations severely deteriorates the numerical stability. To reduce this problem, during the simulations we have increased the frequency of the Green's function recalculation, which we set to be performed whenever the accepted change in the Green's function exceeds some threshold. In the results presented in this letter, the QMC simulations exhibit a satisfactory numerical stability.
It is important to note that, even with an increased frequency of the Green's function recalculation, the computational cost to sample $\hat{\rho}_A$ remains polynomial in $\beta$ and in the volume, as it is in the standard auxiliary field QMC \cite{AF_notes}.

 The second issue regards the eigenvalues and the positive definiteness of the sampled reduced density matrix. While obviously $\hat{\rho}_A$ is a positive-definite operator, QMC simulations do not preserve at every configuration the positive-definiteness of $\hat{\rho}_A$ and hence, a finite QMC sample may result in a non-positive definite $\hat{\rho}_A$. In other words, although the sampled eigenvalues of $\hat{\rho}_A$ are positive within error bars, their statistical fluctuations can be large enough so as to fluctuate between positive and negative values. If this is the case, the computation of the logarithm of $\hat{\rho}_A$ is ill-defined. This problem is particularly relevant for the smallest eigenvalue of $\hat{\rho}_A$.
 It should be noted that presence of small eigenvalues is in fact a fundamental obstruction in the calculation of the entanglement Hamiltonian of a given model, found also, e.g., in the case of a free fermionic chain \cite{Peschel-04}.
  Although we are not able to give a physical meaning to the smallest eigenvalue, we observed that this issue is more severe in the computation of the two-body term, rather than the one-body term. To circumvent this problem, one needs to collect a sufficiently large number of MC samples, so as to reduce the statistical fluctuations until the positive-definiteness of the sampled $\hat{\rho}_A$ is ensured. It is important to observe that the normalization of the eigenvalues of $\hat{\rho}_A$ is strictly related to the probability measure used to sample it. It follows from Eq.~(\ref{rhoexpansion}) that the measure $\widetilde{P}(\{s\})$ fixes the normalization such that in the 0-particles sector the eigenvalue of $\hat{\rho}_A$ is 1 or, equivalently, it fixes the $\text{const}$ term in Eq.~(\ref{HEexpansion}) to 0.
 Under the usual normalization $\Tr \hat{\rho}_A=1$, it follows from Eq.~(\ref{rhoAF}) (where such a normalization holds) that the 0-particles eigenvalue is given by the expectation value of $\det\left(\id - G_A(\{s\})\right)$ on the canonical QMC measure $P(\{s\})$. For all cases considered here, it turns out that such an eigenvalue is much smaller than $1$, so that the measure $\widetilde{P}(\{s\})$ effectively normalizes $\hat{\rho}_A$ magnifying its eigenvalues.

In summary, the QMC sampling used in this work significantly ameliorate the fundamental problem of small eigenvalues in $\hat{\rho}_A$, at the price of increasing numerical instabilities.
It is conceivable that a different probability measure may mitigate the aforementioned problems, ideally by providing bigger eigenvalues of $\hat{\rho}_A$ and, at the same time, controlled statistical fluctuations.
While the above issues are general, their severity is model-dependent, as our results show.

\subsection{Expansion of the reduced density matrix}
\label{sm:expansion}
In this section we prove Eq.~(\ref{rhoexpansion}). As in the main text, we assume an implicit summation over repeated indexes, unless otherwise stated. The exponential appearing on the left-hand side of Eq.~(\ref{rhoexpansion}) resembles an operator creating a Gaussian state. However, it is important to notice that the matrix $h(\{s\})$ is generically nonhermitian, since hermiticity of the model is broken by the QMC algorithm at the level of single realizations of $\{s\}$, and recovered only after the average over the QMC configurations. Still, one can safely assume that $h(\{s\})$ can be diagonalized, since the set of diagonalizable matrices is dense. For the sake of simplicity, in the following we neglect the dependence of the various matrices on $\{s\}$. Let $U$ the matrix such that
\begin{equation}
  h = U^{-1}DU, \qquad D={\rm diag}(\lambda_1,\ldots,\lambda_N),
  \label{Ddef}
\end{equation}
where $\lambda_i$ are the eigenvalues of $h$ (not necessarily real).  We define the following operators:
\begin{equation}
\begin{split}
\gdag{} &\equiv \adag{} U^{-1} \\
\fd{} &\equiv U\ad{}\\
\mop{i} &\equiv \gdag{i}\fd{i},
\end{split}
\label{gfn_def}
\end{equation}
where $\ad{}$ (respectively, $\adag{}$) is the column (respectively, row) vector of operators $\{\ad{i}\}$ (respectively, $\{\adag{i}\}$), and no summation over the index $i$ is implied in the definition of $\mop{i}$. Using Eq.~(\ref{gfn_def}), the term in the exponential of Eq.~(\ref{rhoexpansion}) can be written as
\begin{equation}
\adag{} h \ad{} = \adag{} U^{-1}D U \ad{} = \gdag{} D \fd{} = \lambda_i \mop{i}.
\label{aha_n}
\end{equation}
If $h$ is hermitian, one can find $U$ such that $U^{-1}=U^\dag$ and $\fd{}=\gd{}$, but in general this does not hold and the transformation is not canonical. Nevertheless, the useful (anti-)commutation relations are still valid. In particular:  
\begin{align}
&\left\{\fd{i},\gdag{j}\right\} = \left\{U_{il}\ad{l}, \adag{m}U^{-1}_{mj}\right\}= U_{il}U^{-1}_{lj} = \delta_{ij}
\label{anti_fg}\\
&\left\{\gdag{i},\gdag{j}\right\} = \left\{\adag{l}U^{-1}_{li}, \adag{m}U^{-1}_{mj}\right\} = 0
\label{anti_gg}\\
&\left\{\fd{i}, \fd{j}\right\} = \left\{U_{il}\ad{l},U_{jm}\ad{m}\right\}=0
\label{anti_ff}
\end{align}
Eqs.~(\ref{anti_gg}) and (\ref{anti_ff}) imply also that
\begin{equation}
(\gdag{i})^2 = 0, \qquad (\fd{i})^2 = 0.
\label{gf2}
\end{equation}
By repeatedly using the anticommutation relations given in Eqs.~(\ref{anti_fg})-(\ref{anti_ff}), it is easy to show that
\begin{equation}
  \mop{i}\mop{j} =\mop{j}\mop{i} + \delta_{ij}\left(-\gdag{j} \fd{i} +\gdag{i} \fd{j}\right),
\label{ninj}
\end{equation}
where the right-hand side is not meant to be summed over the index $j$. In Eq.~(\ref{ninj}), the last term vanishes because either $i\ne j$ so that $\delta_{ij}$ in front vanishes, or $i=j$ and the term in the parenthesis vanishes. We conclude that
\begin{equation}
[\mop{i}, \mop{j}]=0.
\label{comm_m}
\end{equation}
We also need to compute $\mop{i}^2$. By employing Eq.~(\ref{anti_fg}) we have
\begin{equation}
  \mop{i}^2 = \gdag{i}\fd{i}\gdag{i}\fd{i}=\gdag{i}\left(1-\gdag{i}\fd{i}\right)\fd{i}=\gdag{i}\fd{i}-\gdag{i}\gdag{i}\fd{i}\fd{i},
  \label{n2_intermediate}
\end{equation}
where no summation over $i$ is intended. The last term on the right-hand side of Eq.~(\ref{n2_intermediate}) vanishes because of Eq.~(\ref{gf2}), thus
\begin{equation}
\mop{i}^2 = \mop{i}.
\label{n2}
\end{equation}
The results obtained so far allow to compute the exponential appearing on the left-hand side of Eq.~(\ref{rhoexpansion}). First, we observe that, due to Eq.~(\ref{aha_n}) and Eq.~(\ref{comm_m}), $\exp\{-\adag{}h\ad{}\}$ can be expressed as a product of mutually commuting terms $\exp\{-\lambda_i\mop{i}\}$ (with no summation over $i$). Each of such exponentials can be computed as
\begin{equation}
  \begin{split}
    e^{-\lambda_i \mop{i}}&=1 + \sum_{k=1}^\infty \frac{1}{k!}(-\lambda_i)^k \mop{i}^k= 1 + \sum_{k=1}^\infty \frac{1}{k!}(-\lambda_i)^k\mop{i}\\
    &= 1 + (e^{-\lambda_i}-1)\mop{i},
  \end{split}
\label{expni}
\end{equation}
where no summation over $i$ is implied, and we have used Eq.~(\ref{n2}). Using Eq.~(\ref{aha_n}) and Eq.~(\ref{expni}), we have
\begin{equation}
e^{-\adag{} h\ad{}} = \prod_i\left[1 + (e^{-\lambda_i}-1)\mop{i}\right].
\label{expni_prod}
\end{equation}
In Eq.~(\ref{expni_prod}) each term in the product commutes with each other [see Eq.~(\ref{comm_m})], therefore by developing the product, we obtain
\begin{equation}
\begin{split}
  e^{-\adag{} h\ad{}} &= 1 + \sum_k (e^{-\lambda_i}-1)\mop{k} \\
  &\qquad + \sum_{i < j}(e^{-\lambda_i}-1)\mop{i}(e^{-\lambda_j}-1)\mop{j}+\ldots\\
  &=1+\sum_{n\ge 1} \sum_{k_1<\ldots <k_n}(e^{-\lambda_{k_1}}-1)\mop{{k_1}}\cdot\ldots \\
  &\qquad\qquad\qquad\qquad\quad\cdot(e^{-\lambda_{k_n}}-1)\mop{{k_n}}\\
  &=1+\sum_{n\ge 1}\frac{1}{n!}\sum_{\substack{k_1,\ldots ,k_n\\k_p\ne k_q}}\prod_{l=1}^n (e^{-\lambda_{k_l}}-1)\mop{{k_l}}
\end{split}
\label{expni_sum}
\end{equation}
In Eq.~(\ref{expni_sum}) the right-hand side contains a product of $\mop{i}$ operators which can be written in terms of $\gdag{j}$ and $\fd{j}$ using Eq.~(\ref{gfn_def}):
\begin{equation}
  \prod_{l=1}^n\mop{{k_l}}=\gdag{{k_1}}\fd{{k_1}}\cdot\ldots\cdot\gdag{{k_n}}\fd{{k_n}}
  \label{mi_gf}
\end{equation}
Since in each term of the sum of Eq.(\ref{expni_sum}) the indexes $k_1\ldots k_n$ are different with each other, the operators $g^\dag_{k_i}$, $f_{k_j}$ on the right-hand side of Eq.~(\ref{mi_gf}) all anticommute with each other [see Eqs.~(\ref{anti_fg})-(\ref{anti_ff})]. They can be brought in a normal-order form by bringing the $\gdag{i}$ operators on the left and the $\fd{j}$ operator on the right as $\gdag{{k_1}}\cdot\ldots\cdot \gdag{{k_n}}\fd{{k_n}}\cdot\ldots\cdot \fd{{k_1}}$. For $n\ge 2$, we need $1$ exchange for moving $\gdag{k_2}$ to the left, $2$ for $\gdag{k_2}$, etc.., for a total of $1+\ldots (n-1)=n(n-1)/2$ single permutations of $\gdag{i}$ operators. 
The same number of exchanges are needed to reorder $\fd{k_1}\cdot\ldots\cdot \fd{k_n}$ to $\fd{k_n}\cdot\ldots\cdot \fd{k_1}$, such that the parity of the complete reordering is $1$ and we have
\begin{equation}
  \gdag{{k_1}}\fd{{k_1}}\cdot\ldots\cdot\gdag{{k_n}}\fd{{k_n}}=+\gdag{{k_1}}\cdot\ldots\cdot \gdag{{k_n}}\fd{{k_n}}\cdot\ldots\cdot \fd{{k_1}}.
  \label{reordering}
\end{equation}
By using Eq.~(\ref{reordering}) the $n-$term in the sum of Eq.~(\ref{expni_sum}) can be written as
\begin{equation}
\begin{split}
  \sum_{\substack{k_1,\ldots ,k_n\\k_p\ne k_q}}&\gdag{{k_1}}\cdot\ldots\cdot \gdag{{k_n}}\fd{{k_n}}\cdot\ldots\cdot \fd{{k_1}}\cdot\\
  &(e^{-\lambda_{k_1}}-1)(e^{-\lambda_{k_2}}-1)\cdot\ldots\cdot(e^{-\lambda_{k_n}}-1).
\end{split}
\end{equation}
We can now relax the constraint $k_p\ne k_p$ because, if two indexes are equal, due to Eq.~(\ref{gf2}) the product vanishes. Inserting back Eq.~(\ref{gfn_def}), we obtain
\begin{equation}
\begin{split}
  \sum_{k_1,\ldots ,k_n}\sum_{\substack{i_1,\ldots, i_n \\ j_1,\ldots,j_n}}&\adag{{i_1}}U^{-1}_{i_1,k_1}\cdot\ldots\cdot \adag{{i_n}}U^{-1}_{i_n,k_n} \cdot\\
  &U_{k_n,j_n}\ad{{j_n}}\cdot\ldots\cdot U_{k_1,j_1}\ad{{j_1}}\cdot\\
  &(e^{-\lambda_{k_1}}-1)\cdot\ldots\cdot (e^{-\lambda_{k_n}}-1)
\end{split}
\label{expni_sum_intermediate}
\end{equation}
We can now carry out the sum over $k_1,\ldots,k_n$. Using Eq.~(\ref{Ddef}), the sum over $k_p$ is
\begin{equation}
\sum_{k_p}U^{-1}_{i_p,k_p}(e^{-\lambda_{k_p}}-1)U_{k_p,j_p}=\left(e^{-h}-\id\right)_{i_p,j_p}
\label{single_kp_sum}
\end{equation}
Inserting Eq.~(\ref{single_kp_sum}) in Eq.~(\ref{expni_sum_intermediate}), and the result in Eq.~(\ref{expni_sum}), we finally obtain the expansion of $e^{-\adag{} h\ad{}}$ to any order
\begin{equation}
\begin{split}
e^{-\adag{} h\ad{}} =& \\
  1+ \sum_{n\ge 1}\frac{1}{n!}\sum_{\substack{i_1,\ldots, i_n \\ j_1,\ldots,j_n}}&\adag{{i_1}}\cdot\ldots\cdot \adag{{i_n}}\cdot\\
  &\left(e^{-h}-\id\right)_{i_1,j_1}\cdot\ldots\cdot \left(e^{-h}-\id\right)_{i_n,j_n}\cdot\\
  &\ad{{j_n}}\cdot\ldots\cdot \ad{{j_1}}.
\end{split}
\label{exp_allorders}
\end{equation}
The truncation of the sum to $n=2$ reproduces Eq.~(\ref{rhoexpansion}). In the sum of Eq.~(\ref{exp_allorders}) only the terms which are antisymmetric in the creation indexes $i_1,\ldots,i_n$ and in the annihilation indexes $j_1,\ldots,j_n$ contribute to the sum. It is therefore convenient to accordingly antisymmetrize the product of matrices.

\subsection{Computation of $\hat{H}_E$}
\label{sm:log}
In this section we illustrate how to compute the negative logarithm of $\hat{\rho}_A$, i.e. the entanglement Hamiltonian, in an expansion of normal-ordered many-body operators, as in Eq.~(\ref{HEexpansion}). As is evident from Eq.~(\ref{rhoexpansion}) and Eq.~(\ref{exp_allorders}), $\hat{\rho}_A$ conserves the particle number. This statement generically holds when the Hamiltonian conserves the particle number $\hat{N}$ and the bipartition is such that the operator $\hat{N}$ is split into commuting number operators for the subsystem $A$ and $B$, that is, if $\hat{N}=\hat{N}_A+\hat{N}_B$, with $[\hat{N}_A, \hat{N}_B] = 0$.

The Hilbert space ${\cal H}_A$ for the subsystem $A$ can be decomposed into a direct sum of subspaces with a fixed particle number $n$ as ${\cal H}_A = {\cal H}_A^{(0)} \oplus{\cal H}_A^{(1)}\oplus \ldots {\cal H}_A^{(n)} \oplus \ldots$. A orthonormal base for the $n-$particles subspace ${\cal H}_A^{(n)}$ can be formed using the free-particles states $|v_i^{(n)}\>=\adag{i_1}\cdot\ldots\cdot\adag{i_n}|0\>$. Choosing such bases for ${\cal H}_A^{(n)}$, we first compute the matrix elements of $\hat{\rho}_A$ in ${\cal H}_A$. Due to the particle-number conservation, the matrix is block diagonal. In the block corresponding to ${\cal H}_A^{(n)}$, the matrix elements of $\hat{\rho}_A$ are of the form
\begin{equation}
  \<v_j^{(n)}|\hat{\rho}_A|v_i^{(n)}\> = \<0|\ad{j_n}\cdot\ldots\cdot\ad{j_1}\hat{\rho}_A\adag{i_1}\cdot\ldots\cdot\adag{i_n}|0\>.
  \label{matrix_el_rho}
\end{equation}
Given the matrix representation $M$ of $\hat{\rho}_A$, the matrix $N\equiv -\log(M)$ is, by definition, the matrix representation of $\hat{H}_E$, in the same base of free-particles states. As for $M$, $N$ is block-diagonal and in the subspaces ${\cal H}_A^{(n)}$ its elements $N_{ij}^{(n)}$ are of the form [compare with Eq.~(\ref{matrix_el_rho})]
\begin{equation}
N_{ij}^{(n)}=\<v_j^{(n)}|\hat{H}_E|v_i^{(n)}\>.
\label{matrix_el_He}
\end{equation}
Since the base is orthonormal, we can write the operator $\hat{H}_E$ as
\begin{equation}
  \begin{split}
    \hat{H}_E &= \sum_n\sum_{ij} N_{ij}^{(n)}|v_i^{(n)}\>\<v_j^{(n)}|\\
    &= \sum_n\sum_{ij} N_{ij}^{(n)}\adag{i_1}\cdot\ldots\cdot\adag{i_n}|0\>\<0|\ad{j_n}\cdot\ldots\cdot\ad{j_1}
  \end{split}
  \label{HE_from_Nij}
\end{equation}
where the first sum is over the subspaces ${\cal H}_A^{(n)}$, and the second sum is over the base elements of ${\cal H}_A^{(n)}$. Finally, the vacuum projector $|0\>\<0|$ can be written as
\begin{equation}
  |0\>\<0| = \prod_k \left(1 - \nop{k}\right)= \prod_k \left(1 - \adag{k}\ad{k}\right).
  \label{vacuum}
\end{equation}
By inserting Eq.~(\ref{vacuum}) into Eq.~(\ref{HE_from_Nij}) we finally obtain
\begin{equation}
  \hat{H}_E= \sum_n\sum_{ij} N_{ij}^{(n)}\adag{i_1}\cdot\ldots\cdot\adag{i_n}\left[\prod_k \left(1 - \adag{k}\ad{k}\right)\right]\ad{j_n}\cdot\ldots\cdot\ad{j_1}.
  \label{HE_N}
\end{equation}
It is important to observe that in Eq.~(\ref{vacuum}) and Eq.~(\ref{HE_N}) the product over the indexes $k$ extends over all the single-particle labels of ${\cal H}_A$, whereas for each $n$ the sum over $i$, $j$ refers to the $n-$particles free states of ${\cal H}_A$. After a normal-reordering, Eq.~(\ref{HE_N}) provides an explicit expression for $\hat{H}_E$.
An analysis program which computes the entanglement Hamiltonian from the QMC measurements along these lines has been developed using the TRIQS library \cite{triqs} for manipulating the fermionic operators, and the Armadillo library \cite{armadillo,armadillo18} for the linear algebra.

Several important simplifications are in order. Inspecting the right-hand side of Eq.~(\ref{HE_N}), we observe that a normal-reordering of a term with particle-number index $n$ results in a sum of normal-ordered $m$-body operators with $m\ge n$. Therefore, if we want to determine $\hat{H}_E$ in a normal-order expansion, as in Eq.~(\ref{HEexpansion}), up to the $m-$body term it is sufficient to truncate the sum on the right-hand side of Eq.~(\ref{HE_N}) to $n\le m$. It follows that we only need to compute matrix elements of $\hat{H}_E$, and accordingly of $\hat{\rho}_A$, in the subspaces ${\cal H}_A^{(n)}$, with $n\le m$ particles. In an expansion of $\hat{\rho}_A$ into normal-ordered many-body terms, a $k-$body term contains $k$ annihilation operators on the right. Upon inserting it in Eq.~(\ref{matrix_el_rho}), it can give a nonzero matrix element only in subspaces with $n\ge k$ particles. Thus, in the computation of matrix elements in the subspaces ${\cal H}_A^{(n)}$, $n\le m$, only the $k-$body terms with $k \le m$ can give a nonzero contribution. All in all, a determination of $\hat{H}_E$ up to the $m-$body terms requires the computation of $\hat{\rho}_A$ up to the $m-$body terms. This justifies the truncation to the two-body terms done in Eq.~(\ref{rhoexpansion}) and Eq.~(\ref{HEexpansion}), when calculating $\hat{H}_E$ up to the two-body terms. For a fixed $L_a$, the computational cost grows approximately exponential in $m$ when $m\ll L_a$, and saturates to a cost exponentially large in $L_a$ when $m\gtrsim L_a$.

An additional interesting simplification occurs for the computation of the one-body term $t_{ij}$ in Eq.~(\ref{HEexpansion}) of the entanglement Hamiltonian. Up to an inessential normalization constant, $\hat{\rho}_A$ is the expectation value of the right-hand side of Eq.~(\ref{rhoexpansion}) on the measure $\widetilde{P}$ [see also Eq.~(\ref{rhoPtilde})]. Its matrix elements in the subspaces ${\cal H}_A^{(0)}$ and ${\cal H}_A^{(1)}$ are
\begin{equation}
  \begin{split}
  M^{(0)} &= \<0|1|0\> = 1,\\
  M^{(1)}_{ij} &= \<0|\ad{i}\left[ 1+\adag{k} \left\<\left(e^{-h(\{s\})}-\id\right)_{kl}\right\>_{\widetilde{P}}\ad{l}\right]\adag{j}|0\>\\
  &=\left\<\left(e^{-h(\{s\})}\right)_{ij}\right\>_{\widetilde{P}},
  \end{split}
  \label{onebodym}
\end{equation}
where, in agreement with the previous notation, $\<\ldots\>$ indicates the matrix elements of an operator in ${\cal H}_A$, and $\<\ldots\>_{\widetilde{P}}$ the expectation value over the QMC measure $\widetilde{P}$.
Inserting Eq.~(\ref{onebodym}) into Eq.~(\ref{HE_N}) we find, up to the one-body term,
\begin{equation}
  \hat{H}_E=\text{const} -\log\left(M^{(1)}\right)_{ij}\adag{i}\left[\prod_k \left(1 - \adag{k}\ad{k}\right)\right]\ad{j},
  \label{HE_one_intermediate}
\end{equation}
where the $\text{const}$ term arises from the normalization constant of $\hat{\rho}_A$ implicit in Eq.~(\ref{rhoPtilde}). Up to the one-body term, in the right-hand side of Eq.~(\ref{HE_one_intermediate}) we can substitute the product of $1-\nop{k}$ operators with $1$, leading to a simple expression for the one-body coefficients $t_{ij}$ appearing in Eq.~(\ref{HEexpansion})
\begin{equation}
  t_{ij} = \log\left(M^{(1)}\right)_{ij}.
  \label{tij}
\end{equation}
Eq.~(\ref{tij}) may also be proven by computing the Mercator series for the logarithm of the right-hand side of Eq.~(\ref{rhoexpansion}), after taking the expectation value over $\widetilde{P}$.

The whole approach has been in particular tested against the $U=0$ case, where due to the decoupling of the HS fields, the QMC sampling has no statistical fluctuations. Nonetheless, this constitutes a quite nontrivial consistency check: the QMC produces an exact reduced density matrix $\hat{\rho}_A$ up to the one- and two-body terms, where both of them are nonvanishing. Upon taking the logarithm, we obtain an entanglement Hamiltonian with vanishing two-body terms, while the one-body term reproduces the known exact result for free fermions \cite{Peschel-04}.
\putbib
\end{bibunit}
\end{document}